\documentclass[aps,pre,amsmath,amssymb,groupedaddress]{revtex4}
\usepackage{graphicx}
\usepackage{amsmath}
\newcommand{\be}{\begin{equation}}
\newcommand{\ee}{\end{equation}}
\newcommand{\ba}{\begin{array}{l}}
\newcommand{\ea}{\end{array}}
\newcommand{\re}[1]{(\ref{#1})}
\newcommand{\ci}[1]{\cite{#1}}
\newcommand{\banonum}{\begin{eqnarray*}}
\newcommand{\eanonum}{\end{eqnarray*}}
\newcommand{\baa}{\begin{eqnarray}}
\newcommand{\eaa}{\end{eqnarray}}
\newcommand{\bfr}{\begin{flushright}}
\newcommand{\efr}{\end{flushright}}
\newcommand{\bfl}{\begin{flushleft}}
\newcommand{\efl}{\end{flushleft}}
\newcommand{\lab}[1]{\label{#1}}
\begin{document}
\date{}
\title{Kicked Dirac particle in a box}
\author{V.E. Eshniyazov,D.U. Matrasulov,  J.R. Yusupov}
\affiliation{Turin Polytechnic University in Tashkent,\\ 17 Niyazov Str., 100095, Tashkent, Uzbekistan}

\begin{abstract}
We study quantum dynamics of a kicked relativistic spin-half
particle in a one dimensional box. Time-dependence of the average
kinetic energy and evolution of the wave packet are explored. Kicking potential is introduced as the Lorentz-scalar, i.e., through the mass-term in the Dirac equation. It is found that depending on the values of the kicking parameters $E(t)$ can be periodic, monotonically growing and non-periodic function of time.
Particle transport in the system is studied by considering spatio-temporal evolution of the Gaussian wave packet.Splitting of the packet into two symmetric parts and restoration of the profile of the packet is found.
\end{abstract}
\maketitle

\section{Introduction}

Study of particle dynamics in confined systems is of fundamental and
practical importance for variety of problems in mesoscopic and
nanoscale physics. This causes monotonically growing interest to the
problem and increasing the number of publications in the literature
. The main difference between the physical properties of bulk and
confined quantum systems is caused by the boundary conditions to be
imposed for a quantum-mechanical wave equations describing their
behavior. In the case of a bulk system these conditions are given in
whole space, while for a confined system the boundary conditions
should be imposed in spatially finite domains. This leads to
considerable difference in the macroscopic properties of the bulk
and confined systems making the latter size- and shape-dependent.
Therefore the role of the quantum confinement in particle transport
and its effects on the macroscopic properties of the system are
among the key problems of nanoscale physics. Earlier particle
dynamics in confined domains was the subject of extensive research
in the context of nonlinear dynamics and quantum chaos theory
 (see, e.g., Refs. \ci{Gutz}-\ci{Richter}).This is mainly done via modeling such systems by so-called billiard geometries \ci{Gutz,Uzy}. Billiards are hard-wall boxes that has become one of the paradigms
in dynamical chaos theory \ci{Gutz,Stoeck}. It was found that
depending on the shape of billiard wall classical dynamics of the
system can be regular, mixed or chaotic. In the case of quantum
systems the effect of billiard size and shape appears in eigenvalues
and eigenfunctions of the Schrodinger equation for which billiard
boundary conditions are imposed. For classically non-integrable
(chaotic) billiards the effect of classical chaos is exhibited in
statistical properties of the energy levels and wave function of the
billiard particle \ci{Gutz,Stoeck}.

Apart from the quantum chaos theory, billiards found application in
nanoscale physics as the models of quantum dots \ci{Nak1,Richter}.
Also, the wave dynamics in a microwave cavity is well described by
quantum billiards\ci{Stoeck}. Confined particle motion appear in
different types of nanoscale and mesocopic systems such as quantum
dots, wells and wires, nanoscale networks, fullerene, CNT, graphene
nanoribbons and many other structures. Most of these systems can be
modeled by  quantum billiards \ci{Richter,Nak1,Imry} or graphs
\ci{Uzy}. In all cases, study of particle dynamics in confined
quantum systems is reduced to solving of quantum mechanical wave
equations with the box(billiard) boundary conditions.

Despite the fact that confined quantum systems are extensively
discussed in the literature, the spectrum of the studied problems is
mainly restricted by considering isolated (unperturbed) and
nonrelativistic systems. However, quantum dynamics in driven
confined systems is of importance for many nanoscale systems, as in
many cases they are subjected to the actions of external static or
time-dependent forces. Another restriction  in the past studies of
confined systems is related to nonrelativistic treatment: Most of
the papers on this issue address the nonrelativistic quantum
dynamics described by the Schrodinger equation.

However, confined particle dynamics in relativistic quantum systems
is relevant to few important topics such as MIT bag model  of
particle physics \ci{Mukherjee}, carbon nanotubes \ci{Ando},
graphene \ci{Brey,Neto} and Majorana fermions \ci{Beenakker} in
condensed matter. All these problems require solving of the Dirac
equation with the box boundary conditions. We note that the Dirac
equation for a particle confined in a box  was earlier treated in
detail by Alonso et.al in \ci{Alonso,Alonso1}, Berry and Mondragon
\ci{Berry}. Different formulations of the stationary Dirac equation
for one dimensional box can be found  also in the
Refs.\ci{Roy,Alberto,Menon}.

It is important to note that unlike to the Schrodinger equation for
box , corresponding Dirac equation encounters with some additional
complications. These difficulties are caused by the fact that for
infinitely square well the Dirac equation cannot be treated as a
single particle equation that is the result of the Klein tunneling
and spontaneous electron-positron pair creation \ci{Bagrov,Greiner}.
To avoid such complication, in the Ref.\ci{Berry} the authors
considered the situation when confinement is caused by a
Lorentz-scalar potential, i.e. by a potential coming in the mass
term. Such a choice of confinement is often used in MIT bag model
\ci{MIT} and the potential models of hadrons \ci{Mukherjee}. Another
way to avoid this complication is to impose box boundary conditions
in such a way that they provide zero-current and probability density
at the box walls. In the Ref. \ci{Alonso} the types of the box
boundary conditions, providing vanishing current at the box walls
and keeping the Dirac Hamiltonian as self- adjoint are discussed. We
note that in condensed matter physics such boundary conditions are
often used to describe the quasiparticle motion in graphene
nanoribbons \ci{Brey,Neto}.

In this paper, to treat delta-kicked relativistic particle dynamics
confined in a one-dimensional box, we will use the boundary
conditions formulated in the Ref. \ci{Alonso}. The kicking potential
is considered as to be Lorentz scalar i.e. coming through the
mass-term.

This paper is organized as follows. In the next section, following
the Ref. \ci{Alonso}, we briefly recall the problem of stationary
Dirac equation for one dimensional box. In section 3 we treat
time-dependent Dirac equation with delta-kicking potential with the
box boundary conditions. Section 4 presents some results and their
analysis. Finally, section 5 presented some concluding remarks.

\section{Dirac particle in a one dimensional box}

Before starting treatment of the kicked Dirac particle dynamics, we
briefly recall the description of the one dimensional stationary Dirac equation
in a box following the Ref.\ci{Alonso}. In nonrelativistic  case the
box boundary conditions for the Schr\"odinger equation are
introduced through the  infinite square well, either by requiring
zero-current at the box walls. In case of the Dirac equation
introducing of infinite well leads to pair production from vacuum.
Therefore the problem cannot be treated within the one-particle
Dirac equation that makes impossible using the infinite well based
description of the particle-in-box system. The second approach, i.e.
requiring zero-current at the boundary can be used, if it does not
lead to the breaking down of the self-adjointness of the problem. In
case of the Schrodinger equation the boundary condition, $\psi=0$
keeps the Schrodinger operator as self-adjoint. However, for Dirac
equation the boundary conditions at the box walls should be
determined carefully.

The stationary Dirac equation for free particle in  a one-dimensional box given on
the interval $(0, L)$ can be written as (in the system of units $m_e
=\hbar =c=1$)
\begin{equation}
H_0\psi=(-i\alpha_x \cdot \frac{d}{dx} + \beta)\psi=E\psi
\lab{dirac_eq}
\end{equation}
where $\alpha_x$ and $\beta$ are the  Dirac matrices. The wave
function, $\psi$ can be written in two component form as
\begin{equation}
\psi=\left( \begin{array}{c}  \phi \\ \chi \end {array} \right)
\end{equation}
where large, $\phi$ and small, $\chi$ components are  also
two-component semi-spinors:
$$
\phi=\left( \begin{array}{c}  \phi_1 \\ \phi_2 \end {array}
\right)\;\;\;\chi=\left( \begin{array}{c}  \chi_1 \\ \chi_2 \end
{array} \right)
$$
respectively.

The system of first order differential equations \re{dirac_eq} can
be reduced to second order, Helmholtz-type equation by eliminating
one of the components:
\begin{equation}
\left( \frac{d^2}{dx^2}+k^2 \right) \phi_i=0\;\;\; i=1,2
\lab{large_comp1}
\end{equation}
Here
\begin{equation}
k=[E^2- 1]^{1/2}. \lab{eigenvalue}
\end{equation}
The small and large components are related to each other through the
expression
\begin{equation}
\left( \begin{array}{c}  \chi_1 \\ \chi_2 \end {array}
\right)=\frac{-i}{E+1}\left( \begin{array}{cc} 0 & \frac{d}{dx} \\
\frac{d}{dx}  & 0  \end {array} \right) \left( \begin{array}{c}
\phi_1 \\ \phi_2 \end {array} \right) \lab{comps_rel}
\end{equation}

Then one of the positive energy solutions can be obtained by taking
$\phi_2=0$ and therefore $\chi_1=0$. Thus the general solution for
$\phi_1$ can be written as
\begin{equation}
\phi_1=A_1 e^{ikx}+B_1e^{-ikx}
\end{equation}
where$A_1$ and $B_1$ are complex constants. For $\chi_2$ one can
obtain
\begin{equation}
\chi_2=\frac{k}{E+ 1}\left( A_1 e^{ikx}-B_1e^{-ikx} \right).
\end{equation}

Now let us  come back to one-dimensional box problem. Since large
and small components are related through Eq.\re{comps_rel}, it is
enough to impose boundary conditions for one of them only, for
example, for large component \ci{Alonso}:
\begin{equation}
\phi_1(0)=\phi_1(L)=0 \lab{condition1}
\end{equation}
Then we the eigenfunctions, corresponding to these boundary
conditions can be written as \ci{Alonso}
\begin{equation}
\psi=2A_1\left( \begin{array}{c}  i \sin(kx) \\ 0 \\ 0 \\
\frac{k}{E+ 1} \cos(kx) \end {array} \right) \lab{ps1}
\end{equation}
 with $k=N \pi/L,N=1,2,....$
It was shown in the Ref. \ci{Alonso} that the boundary conditions
given by Eq. \re{condition1} correspond, in the non-relativistic
limit, to the familiar condition of a vanishing wave function at the
walls of the box: $\phi^{NR}_1(0)=\phi^{NR}_1(L)=0.$

It was also shown in \ci{Alonso} that probability ($\rho$) and
current ($j$) densities defined as
\begin{equation}
\rho= \overline{\phi}_1\phi_1+\overline{\chi}_2\chi_2
\end{equation}
\begin{equation}
j=e\psi^\dagger \alpha_x \psi = ec(\overline{\phi}_1\chi_2
+\overline{\chi}_2^-\phi_1)
\end{equation}
satisfy the following boundary conditions:
\begin{equation}
\rho(0)= \rho(L) \lab{density1}
\end{equation}
\begin{equation}
j(0)= j(L)=0 \lab{current1}
\end{equation}
This implies that the particle is confined inside the box. Dirac
particle confined in a one-dimensional box, whose periodically
driven dynamics we are going to explore in the next section, defined
though the above boundary conditions.

\section{Kicked Dirac particle confined in a one dimensional box}

Now consider the relativistic spin-half particle confined in a box
and interacting with the external delta-kicking potential of the
form
$$
V(x,t) =-\varepsilon\cos(\frac{2\pi x}{\lambda})\sum_l\delta(t -lT)
$$
where $\varepsilon$ and $T$ are the kicking strength and period,
respectively. The dynamics of the system is governed by the
time-dependent Dirac equation which is given as \be i\frac{\partial
\Psi(x,t)}{\partial t} =[-i\alpha\frac{d}{dx} +\beta (1+
V(x,t))]\Psi(x,t), \lab{perturbed1}
 \ee
for which the box boundary conditions are imposed. One should note
that in this equation, to avoid the above mentioned complications
related to becoming of the Dirac equation multi-particle, the
kicking potential is introduced as a Lorentz-scalar, i.e. through
the mass term. Exact solution of Eq.\re{perturbed1} can be obtained
within a single kicking period as in the case of kicked rotor
\ci{Casati} and kicked particle in infinite potential well
\ci{Well}. Indeed, expanding the wave function, $\Psi(x,t)$ in terms
of the complete set of the eigenfunctions given by Eq.\re{ps1} as
$$
\Psi(x,t) = \sum A_n(t) \psi_n(x) \lab{1111}
$$
and inserting  this expansion into  Eq.\re{perturbed1}, by
integrating the obtained equation within one kicking period we have
\be A_n(t+T) =\sum_l A_l(t) V_{ln}e^{-iE_l T}, \lab{evol} \ee where
$$
V_{ln} =\int \psi^\dagger_n(x) e^{i\varepsilon \cos (\frac{2\pi
x}{\lambda})}\psi_l (x)dx
$$
and $E_l$ are defined by Eq. \re{eigenvalue}. Using the relation \be
e^{i\varepsilon \cos x} =\sum_{m=-\infty}^{\infty}
b_m(\varepsilon)e^{im x}, \ee where $b_m(\varepsilon) =i^m
J_m(\varepsilon)$, the matrix elements $V_{ln}$ can be calculated
exactly and analytically. It is clear that the norm conservation in
terms of expansion coefficients, $A_n(t)$ reads as
$$
N(t) = \sum_n |A_n(t)| =1,
$$
that follows from
$$
\int_{0}^{L}|\Psi(x,t)|^2 dx =1\;\; {\rm and}\;
\int_{0}^{L}\psi_m^\dagger(x)\psi_n(x)dx =\delta_{mn}.
$$
In choosing of the initial conditions, i.e. the values of $A_n(0)$ one should use this condition.\\
Having computed $A_n(t)$  we can calculate any dynamical
characteristics of the system such as average kinetic energy or
transport properties. In particular, the average kinetic energy as a
function of time can be written as
$$
E(t) =  \int \Psi^\dagger(x,t)(-i\alpha_x \cdot
\frac{d}{dx})\Psi(x,t)dx ,
$$
 where $A_n(t)$ are given
by Eq.\re{evol}.
\begin{figure}[htb]
\centerline{\includegraphics[width=60mm, angle=-90]{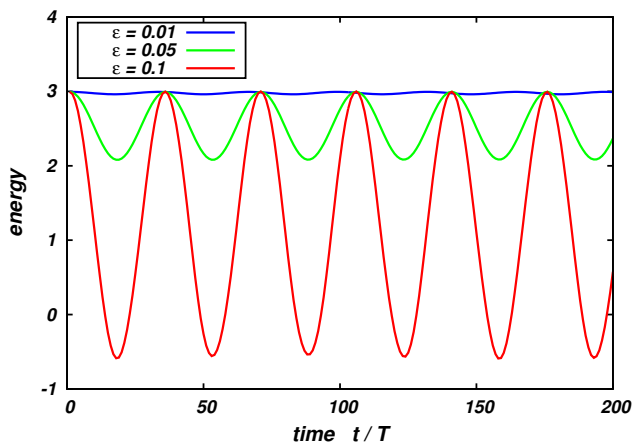}}
\caption{(Color online) Time-dependence of the average kinetic
energy for different values of the kicking strength $\varepsilon =
0.01;\; 0.05;\; {\rm and}\; 0.1\;$ at fixed kicking period $T=0.47$.
} \label{energ1}
\end{figure}
\begin{figure}[htb]
\centerline{\includegraphics[width=60mm, angle=-90]{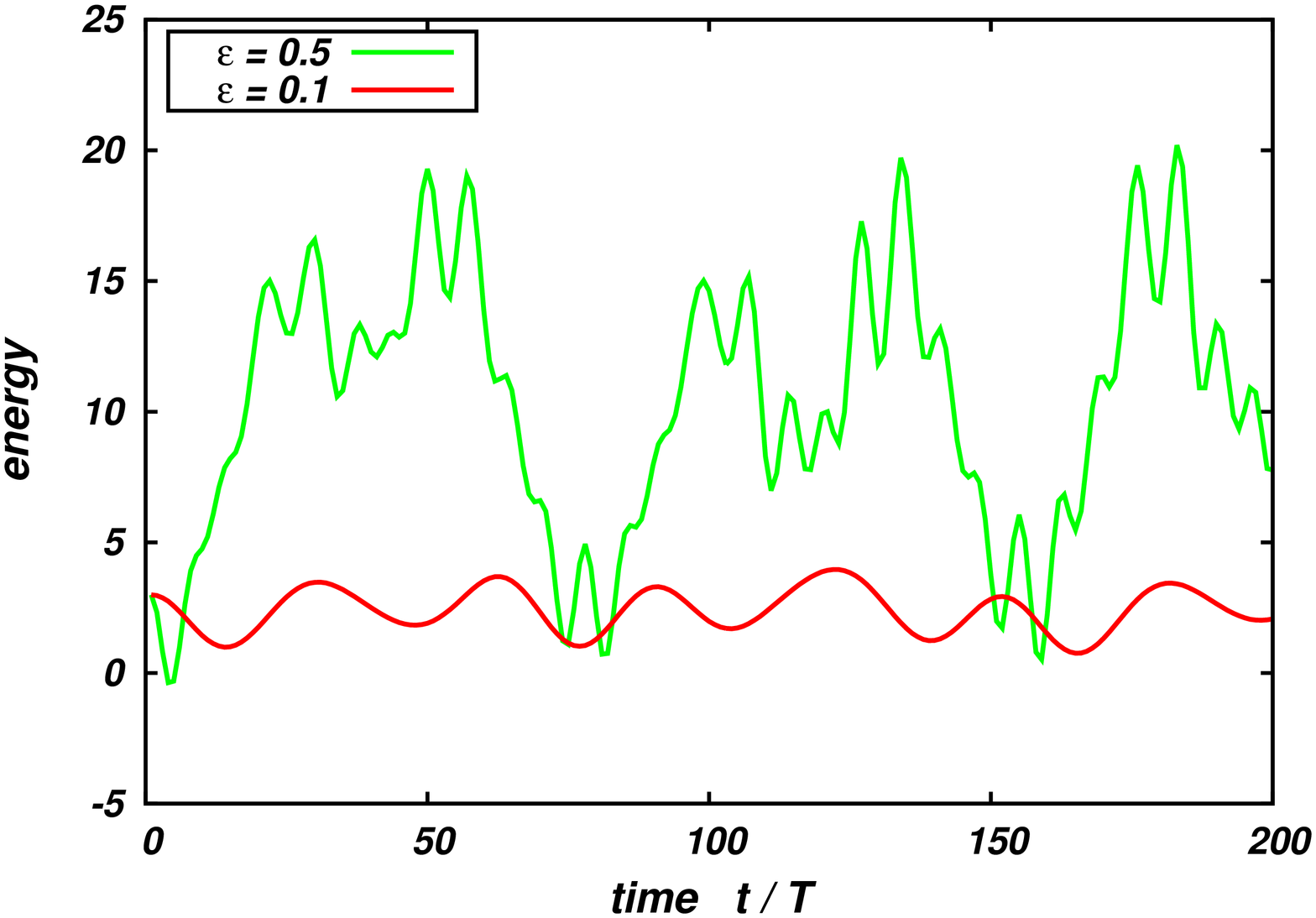}}
\caption{(Color online) Time-dependence of the average kinetic
energy for different values of the kicking strength $(\varepsilon =
0.1;\; {\rm and}\; 0.5\;)$ at fixed kicking period $T=10^{-2}$.}
\label{energ2}
\end{figure}
\begin{figure}[htb]
\centerline{\includegraphics[width=60mm, angle=-90]{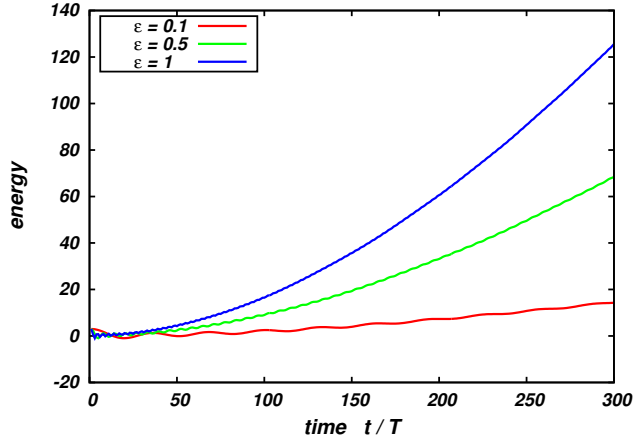}}
\caption{(Color online) Time-dependence of the average kinetic
energy for monotonically growing cases: $(\varepsilon
=0.1;\;0.5;\;1)$ and $T= 10^{-4}$. } \label{energ3}
\end{figure}
\begin{figure}[htb]
\centerline{\includegraphics[width=60mm, angle=-90]{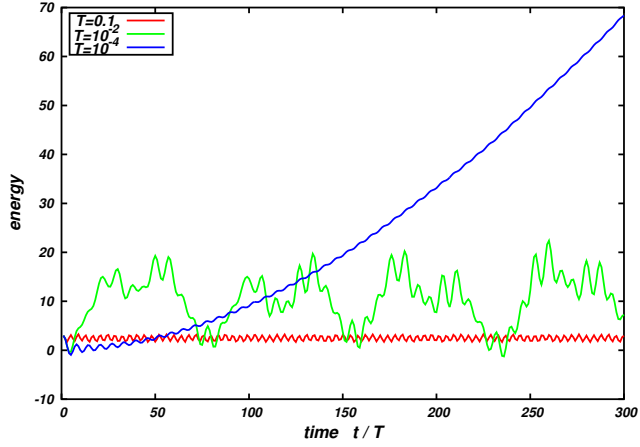}}
\caption{(Color online) Time-dependence of the average kinetic
energy for different kicking periods ($T=0.1;\;10^{-2};\; $ {\rm
and}\; $10^{-4}$) and at fixed value of the kicking strength
($\varepsilon =0.5$).} \label{energ4}
\end{figure}
\begin{figure}[htb]
\centerline{\includegraphics[width=60mm, angle=-90]{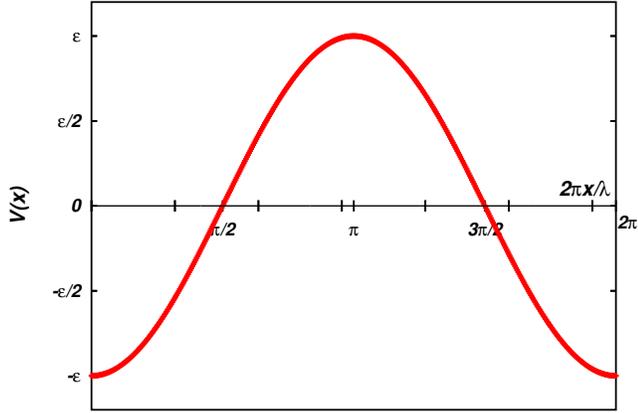}}
\caption{(Color online) Profile of the kicking potential for the
case of $\lambda=L$.}
\end{figure}
\begin{figure}[htb]
\includegraphics[width=60mm, angle=-90]{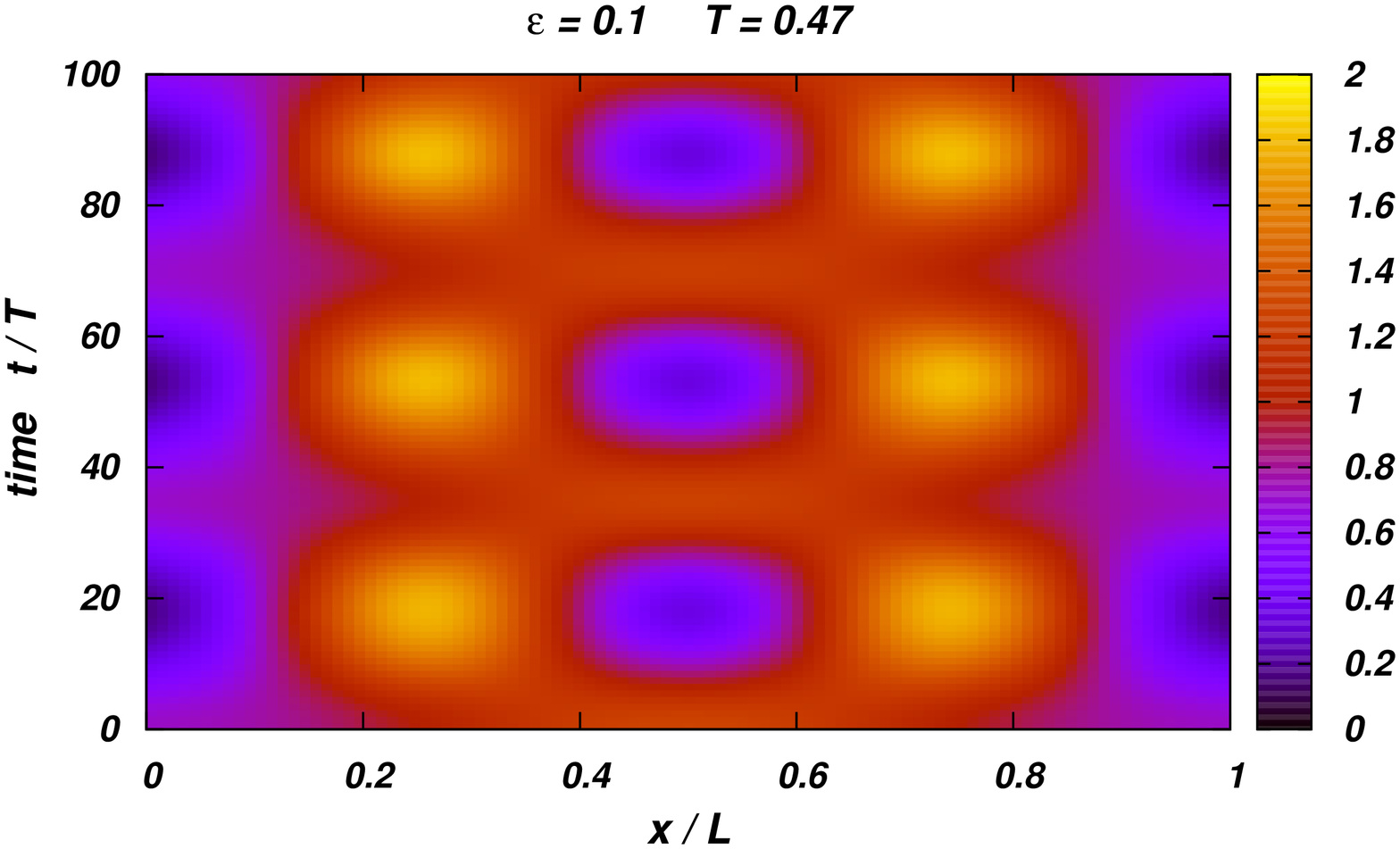}
\includegraphics[width=60mm, angle=-90]{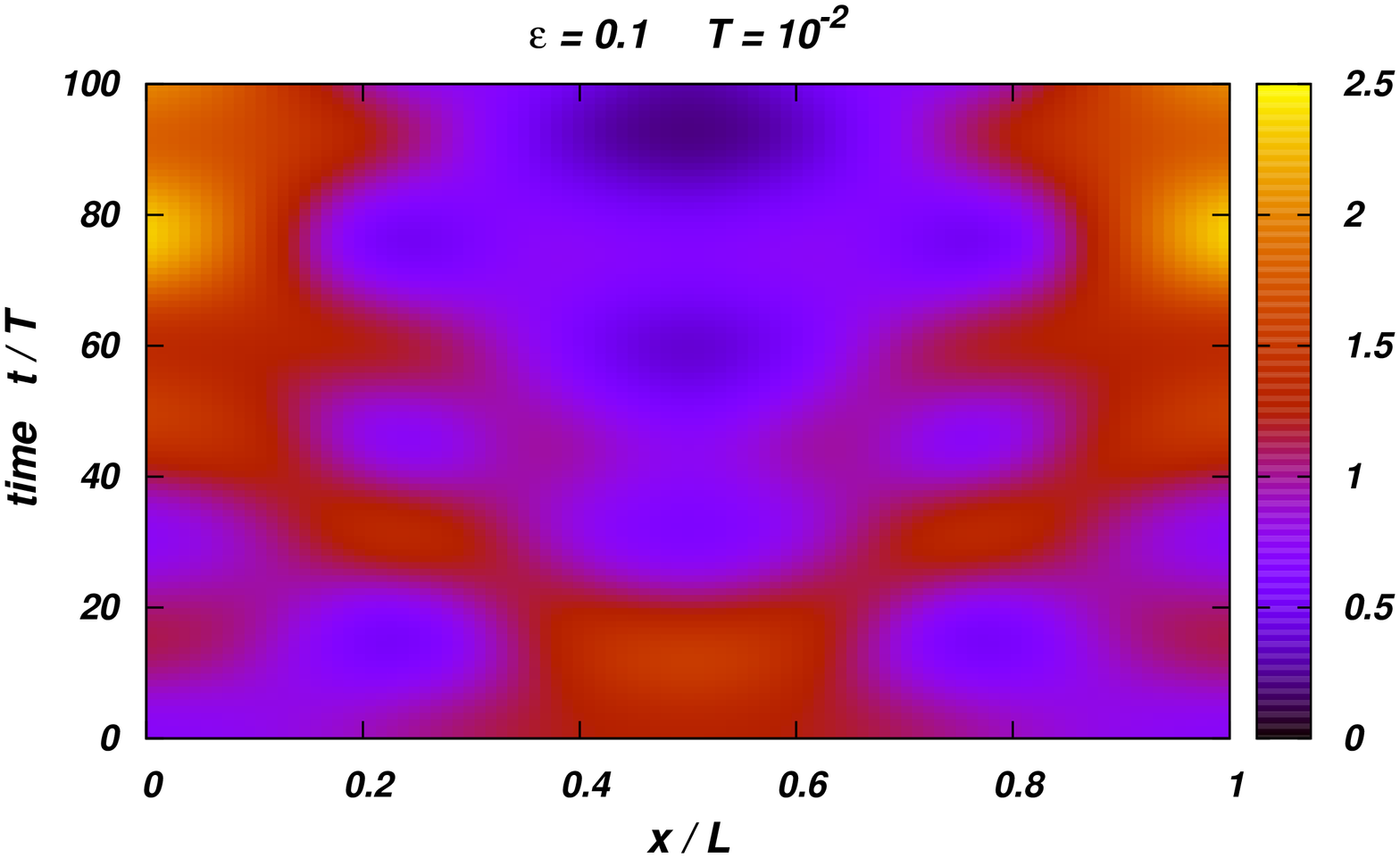}
\centerline{\includegraphics[width=60mm, angle=-90]{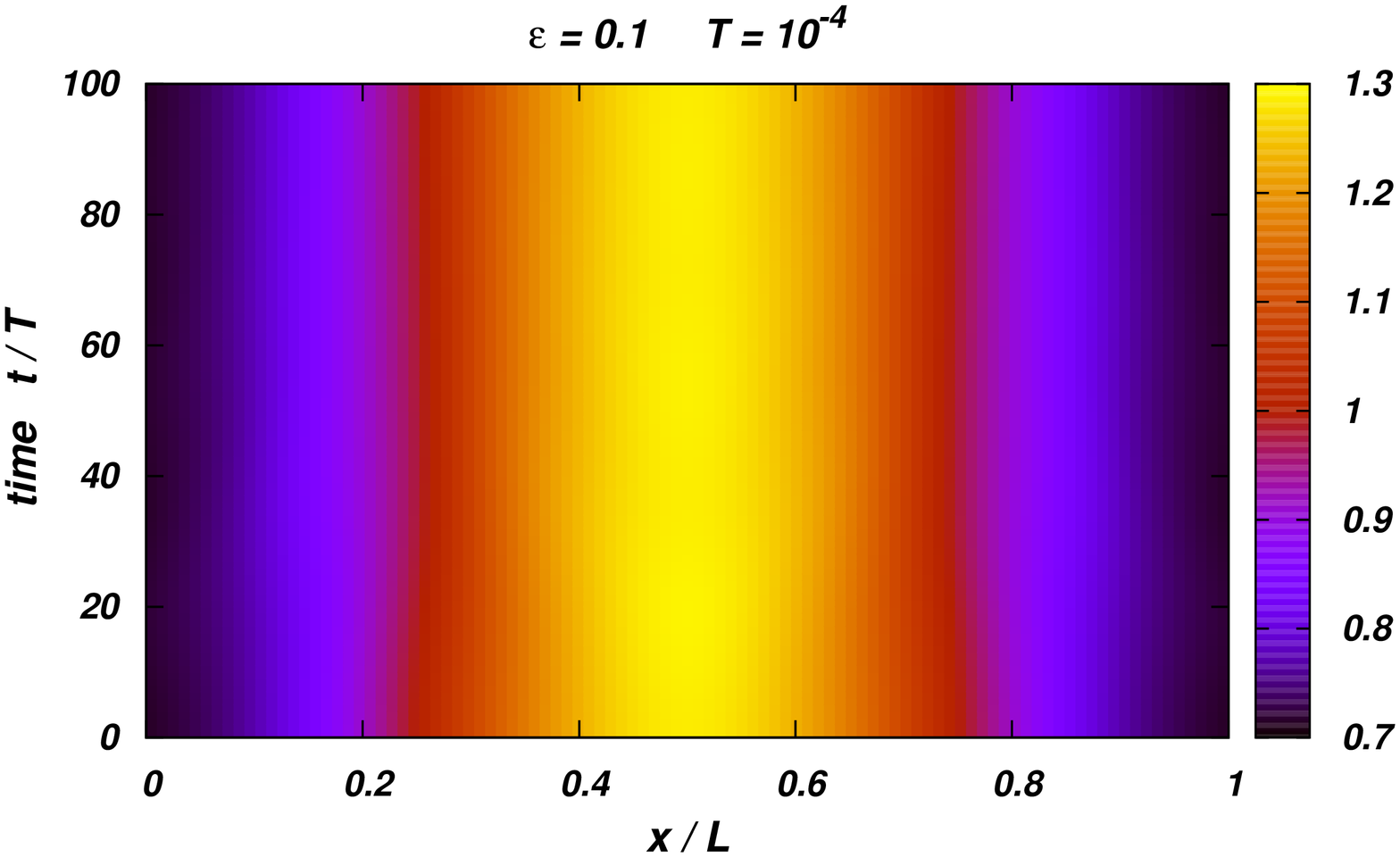}}
\caption{(Color online) Probability density as function of
coordinate and time for the kicking parameters $\varepsilon =0.1$
and $T=0.47$, $T=10^{-2}$ and $T=10^{-4}$.}
\end{figure}
\begin{figure}[htb]
\centerline{\includegraphics[width=60mm,angle=-90]{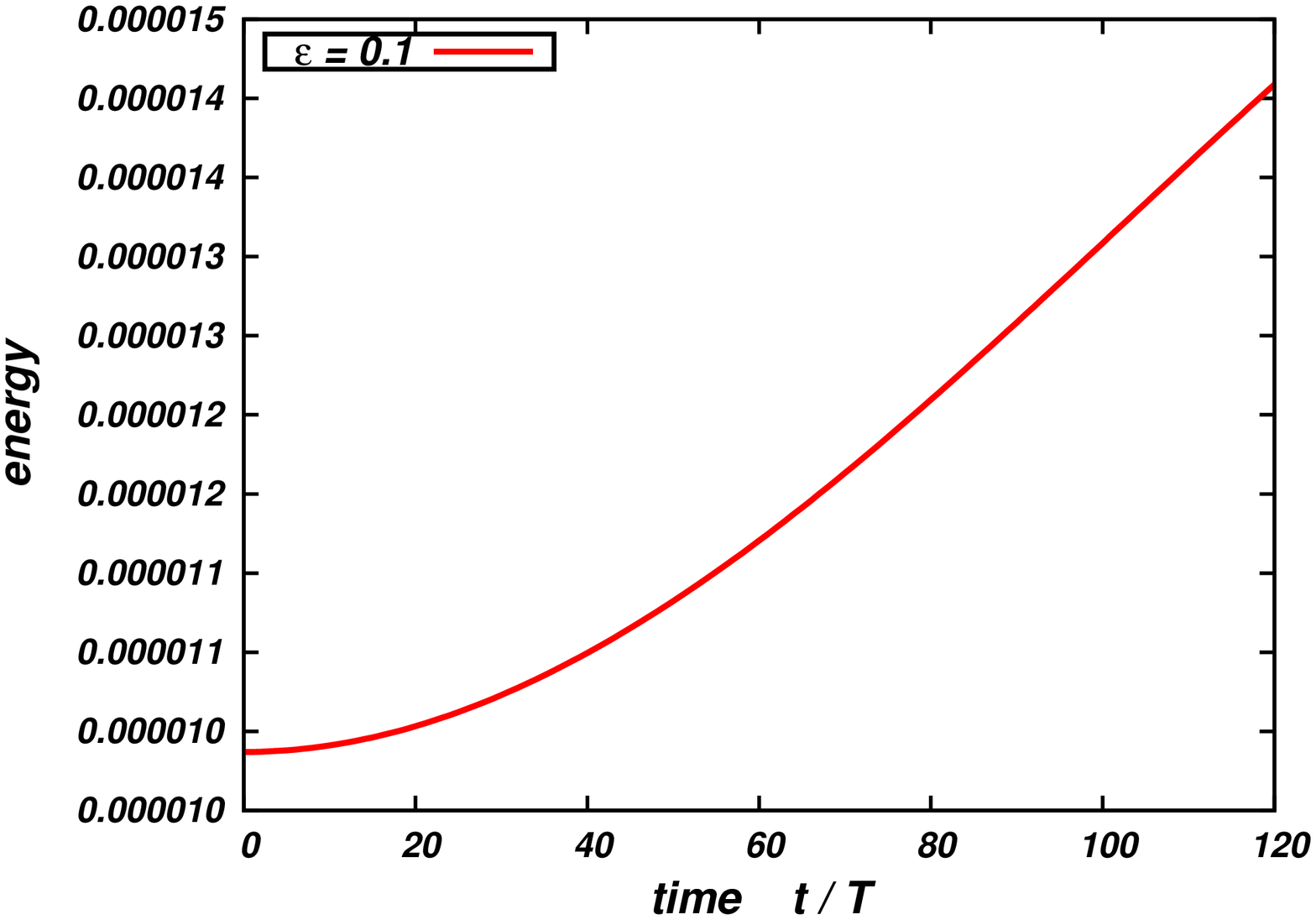}}
\centerline{\includegraphics[width=60mm,angle=-90]{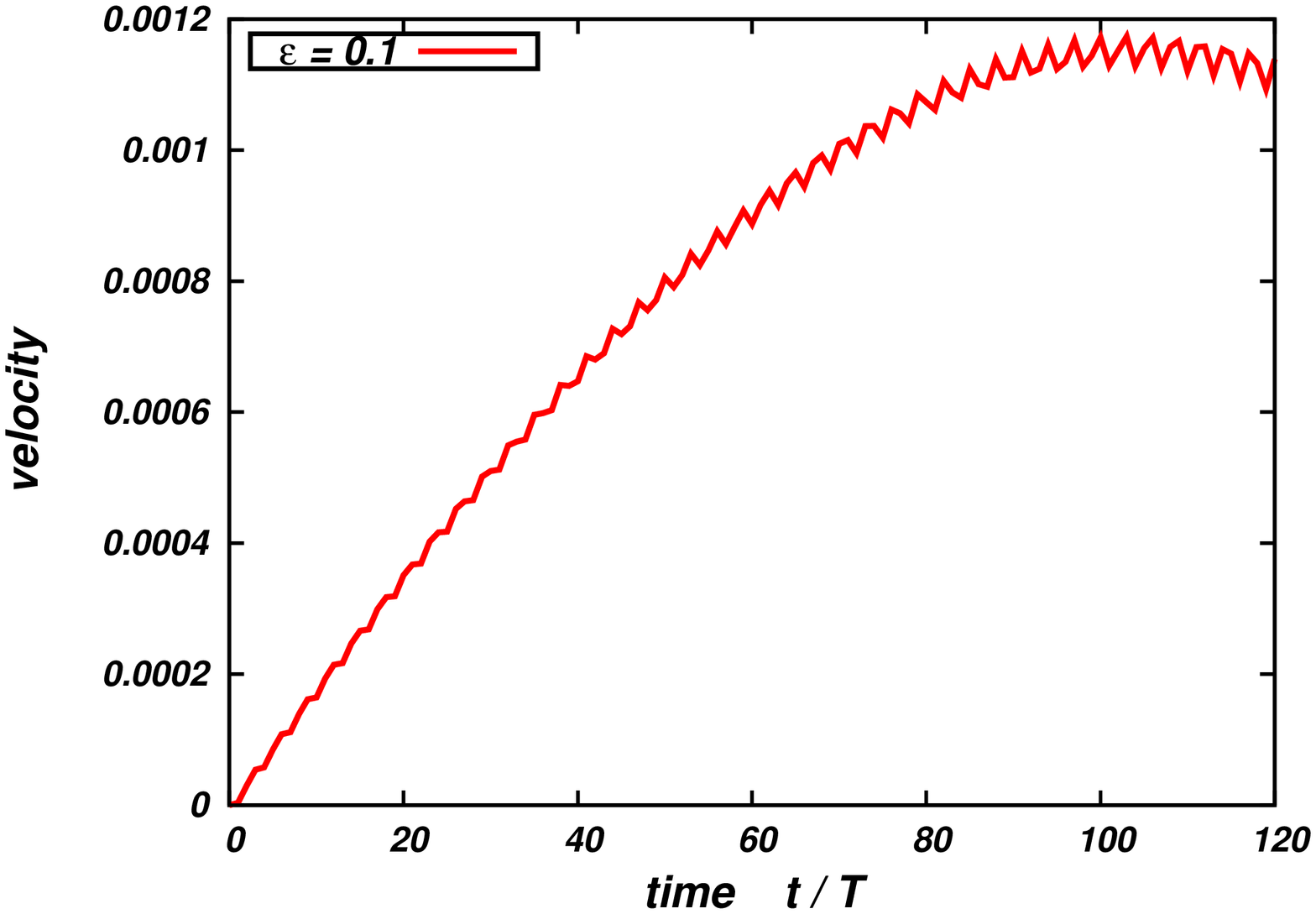}}
\caption{(Color online) Average kinetic energy and average velocity
as a function of time for $\varepsilon=0.1$, $T=100$}
\end{figure}
\begin{figure}[htb]
\centerline{\includegraphics[width=60mm, angle=-90]{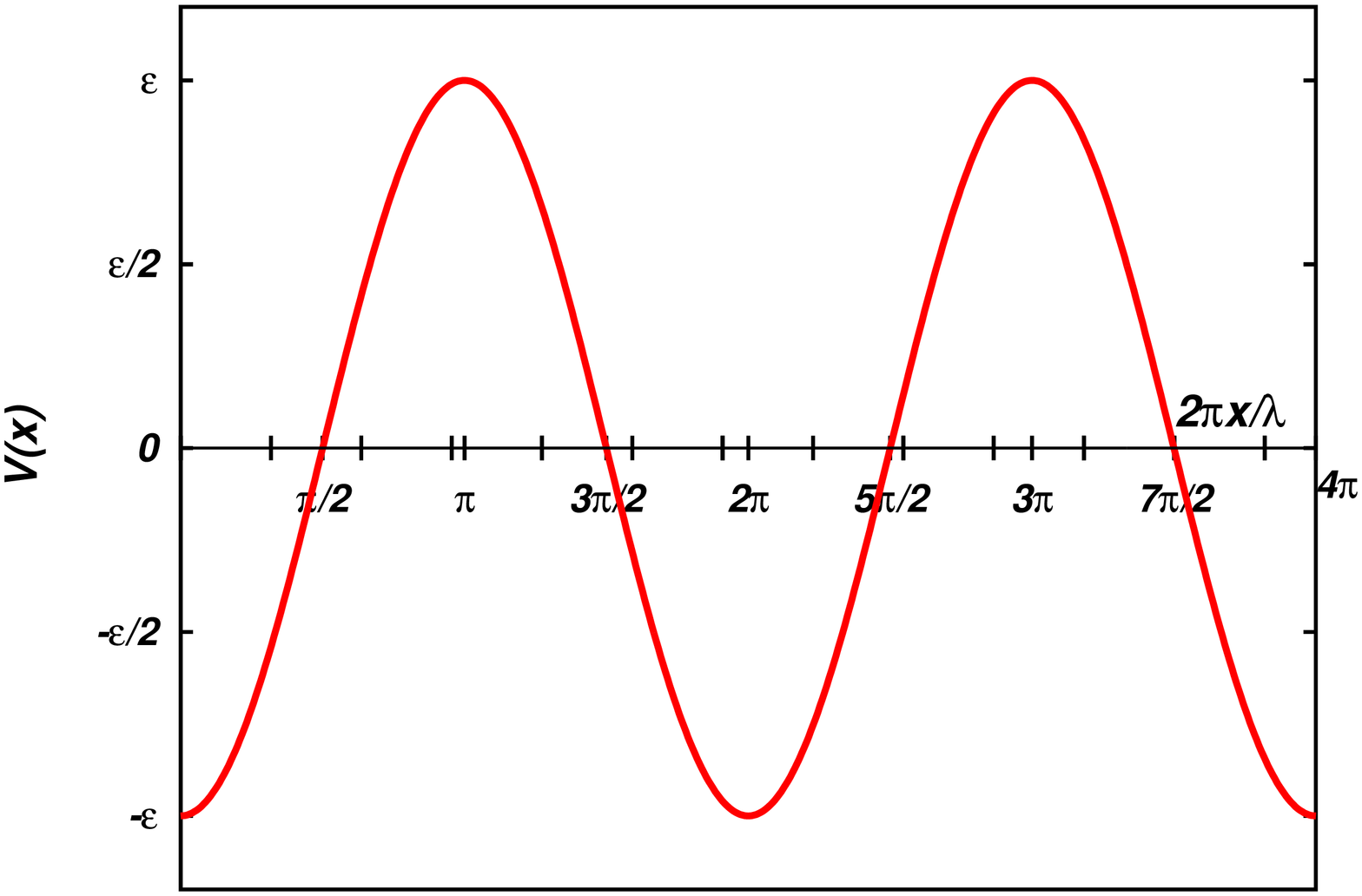}}
\centerline{\includegraphics[width=60mm,angle=-90]{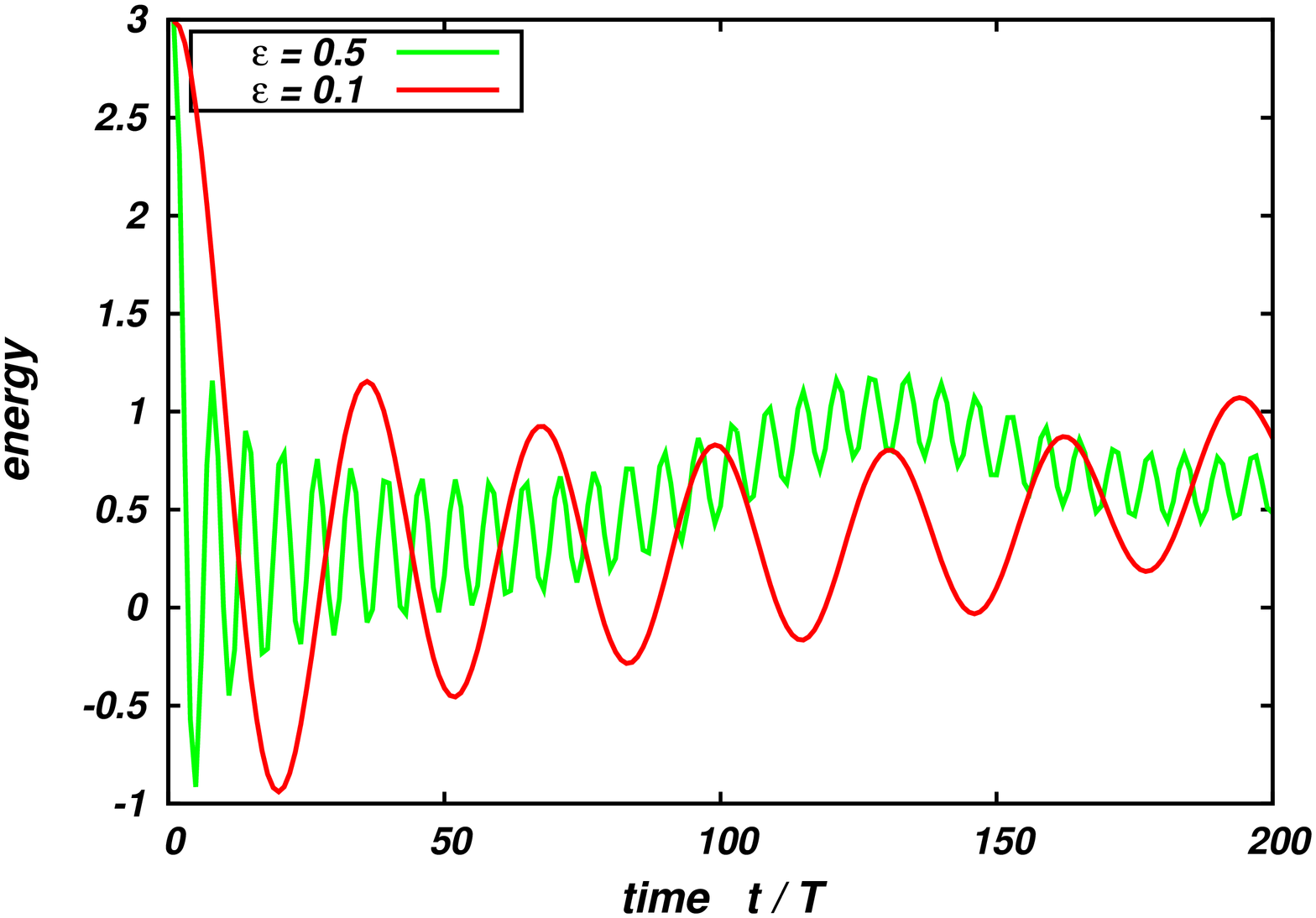}}
\caption{(Color online) Probability density as function of
coordinate and time $\varepsilon =0.1$ and $T=10^{-2}$.}
\label{probdens2}
\end{figure}
For classical periodically driven systems such as kicked rotor or
kicked box, the average kinetic energy grows linearly in time.
However, for quantum counterparts of these systems such a growth is
suppressed \ci{Casati,Izr} which is caused by so-called quantum
localization effect \ci{Izr}. The latter implies that no unbounded
acceleration of a nonrelativistic kicked quantum particle is
possible(except the special cases of quantum resonances \ci{Casati,Izr}). We
are interested to explore the behavior of $E(t)$ in corresponding
relativistic case described by the Dirac equation. Fig. 1 compares
the average kinetic energy as a function of time for different
values of the kicking strength $\varepsilon = 0.01;\; 0.05;\; {\rm
and}\; 0.1\;$ at fixed kicking period $T=0.47$.As is seen from these
plots, $E(t)$ is periodic in time for this set of kicking
parameters. In Fig. 2 $E(t)$ is plotted for $\varepsilon = 0.5$ and
$0.1$ at $T=0.01$. The periodicity in time completely broken in this
case. In Fig.3 monotonically growing cases are presented for the
values of kicking parameters $\varepsilon =0.1;\;0.5;\;1$ and
$T=10^{-4}$. Fig.4 compares the plots of $E(t)$ for different kicking
periods ($T=0.1;\;10^{-2};\; $ {\rm and}\;$10^{-4}$) and at fixed
value of the kicking strength ($\varepsilon =0.5$).

 The difference in behavior of $E(t)$ for different regimes of external kicking
force can be explained by the spatio-temporal localization of the particle with
respect to the profile of the kicking potential. It is clear that
depending on the coordinate, $x$ the kicking potential can be
attractive or repulsive. In Fig. 5 profile of the kicking potential
is plotted for the value of the wavelength $\lambda =1$. If the
motion of the particle is localized in the area where the kicking
potential is positive, gaining of energy by particle and its
acceleration occurs. When particle motion is localized on the area
of the box where the kicking potential is attractive, it loses the
energy. Therefore by tuning the kicking parameters such as
$\varepsilon$, $T$ and $\lambda$ it is possible to achieve tunable
dynamics of a kicked particle in a box. Fig. 6 presents plots of
$\rho(x,t) = |\Psi(x,t)|^2$ for $\varepsilon =0.5$ and $T=10^{-4}$.
As it can be seen, the wave function is localized near the walls and the maximum values are achieved along whole $t-$axis within the
localization band. Fig. 7 presents time dependence of E(t) and
corresponding average velocity as a function of time which is
defined as \ci{Greiner} \be \langle v_x \rangle =-i \int\Psi^\dagger
\alpha_x \Psi dx \ee
 The velocity of a kicked Dirac particle confined in a box does not grow
monotonically and suppressed after some growth.
 The above treatment shows that unlike
to its nonrelativistic counterpart treated in \ci{Well}, the quantum
dynamics of kicked relativistic particle in a box does not depend on
the product $\varepsilon T$, but depends on each kicking parameter
separately. This provides more tools for manipulation by the
dynamics of the system by playing $\varepsilon$, $T$ and $\lambda$.
 Fig. 8 presents plots of $V(x,t)$ for
$\lambda=0.5L$ and $E(t)$ for kicking parameters $T=10^{-4}$ and
$\varepsilon= 0.1$ and $0.5$. If we compare $E(t)$ plotted in Figs.
3 and 8, the strong dependence of the dynamics of the average
kinetic energy on $L/\lambda$ can be observed.

\section{Wave packet evolution}
An important characteristics of the particle transport in quantum
regime is wave packet evolution. Changes of the profile of a packet
in space and time can give useful information about the transport
phenomena. As one of the  most simplest and interesting case one can
treat the evolution of the gaussian wave packet. In the relativistic
case the Gaussian (spinor) wave packet can be written as follows
\ci{Max1}:
\begin{equation}
\Psi(x,0)=\frac{f(x)}{\sqrt{|s_1|^2+|s_2|^2+|s_3|^2+|s_4|^2}}\left\{
\begin{array}{c}  s_1 \\ s_2 \\ s_3 \\ s_4  \end {array} \right\}
\lab{initialwp}
\end{equation}

where $s1,s2,s3$ and $s4$ determine the initial spin polarization
and
$$
f(x)=\frac{1}{d\sqrt{\pi}}\exp\left[ -
\frac{(x-x_0)^2}{2d^2}+iv_0x\right].
$$

Multiplying each side of Eq.\re{initialwp} by $\psi^\ast_m(x)$ and
integrating over $x$ from $0$ to $L$, we obtain expression for the
initial values of the expansion coefficients to used in
Eq.\re{evol}:
\begin{equation}
A_n(0)=\int_0^L \psi^\dagger_n(x)\Psi(x,0) dx .
\end{equation}

 Fig. 9 presents profile of the Gaussian wave packet ( for $d=L/100 $, $x_0=L/2$ and $v_0=0$
 ) at different time moments: $t=0,$\;$20T,$\;$50T$ and $80T$.
 Dispersion of the packet by simultaneous splitting into two
 symmetric parts can be observed for this regime of motion. Such splitting is caused by the existence of the spin of particle.
 In Fig 10. the regime of motion at which restoration of the Gaussian wave packet can be observed, is
 plotted. It is clear that the dynamics of the wave packet in
 relativistic spin-half system is considerably different than that
 in corresponding nonrelativistic system. Such a difference is
 caused by the spin of the system and other relativistic effects.

\section{Conclusion}

Thus we have studied kicked Dirac particle dynamics confined in one dimensional box.
Box boundary conditions for the unperturbed Dirac equation are imposed in such a way that they provide zero-current at
the box walls.

Kicking potential is taken as the Lorentz-scalar, i.e. included into the mass term in the Dirac equation.
Time-dependence of the average kinetic energy is analyzed.
It is found that depending on the values of the kicking parameters $E(t)$ can be periodic, monotonically growing and non-periodic function of $t$.
Such different regimes of motion can be explained by the localization of the particle motion with respect to the profile of the
kicking potential.
In particular, if the particle motion  is always localized in the area where potential is repulsive its gains the
energy and acceleration occurs that corresponds to monotonically growing $E(t)$.

If the localization of the particle in repulsive and attractive areas periodically replace each other $E(t)$ becomes
periodic function of $t$. When the motion of the particle is localized the the area where potential is attractive it looses the energy that leads
to deceleration. Therefore the case when particle is localization in attractive and repulsive areas is non-periodic, the average energy
becomes non-periodic and non-growing function of time.The average velocity as function of time is also analyzed and it is found that its growth as a function of time is suppressed even for monotonically  growing $E(t)$. Particle transport in the system is studied by considering spatio-temporal evolution of the Gaussian wave packet.Splitting of the packet into two symmetric parts and restoration of the profile of the packet is found in this case.

The above results obtained in this paper can be useful for the study of time-dependent particle transport in different
nanoscale systems (e.g. graphene nanoribbons, carbon nanotubes, Majorana wires etc) described by Dirac equation.
Extension of the results for two- and three-dimensional cases is rather trivial.

\begin{figure}[htb]
\centerline{\includegraphics[width=60mm, angle=-90]{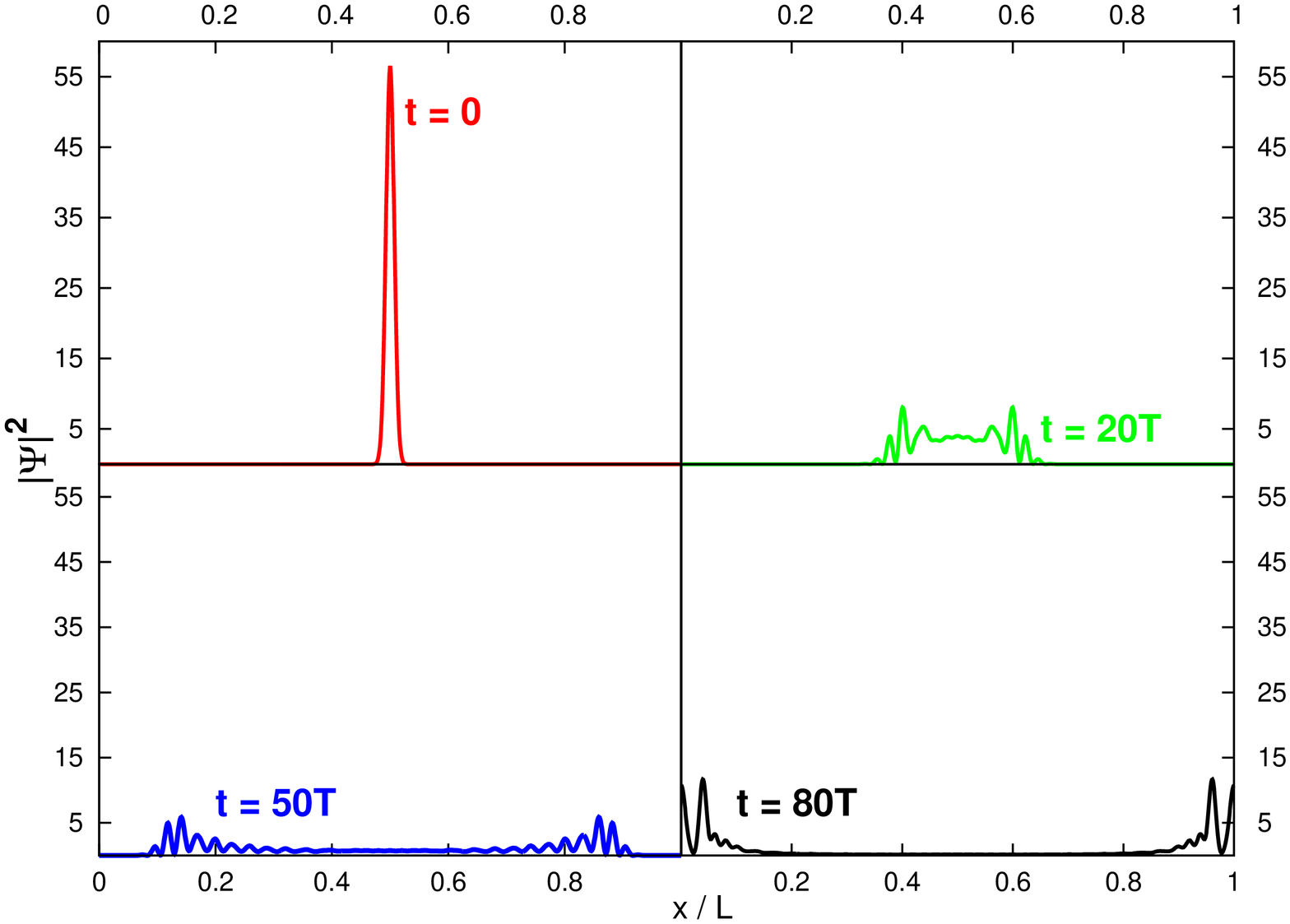}}
\caption{(Color online) Profile of the Gaussian wave packet at
different times for $\varepsilon =1$ and $T=10^{-2}$.}
\end{figure}
\begin{figure}[htb]
\centerline{\includegraphics[width=60mm, angle=-90]{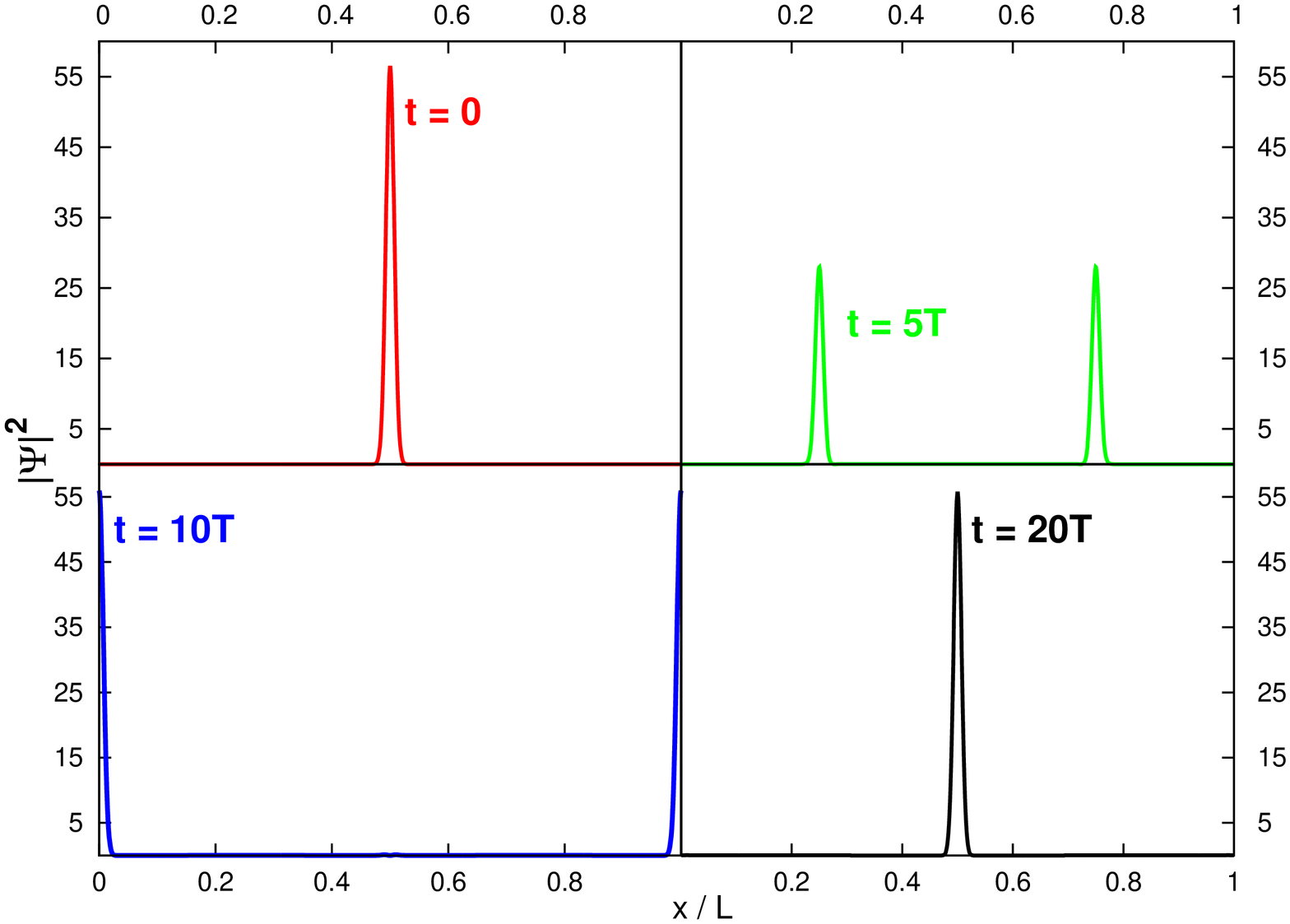}}
\caption{(Color online) Profile of the Gaussian wave packet at
different times for $\varepsilon =1$ and $T=0.25$.}
\end{figure}

\end{document}